%% file: andry.tex
\documentclass[3p,times,twocolumn]{elsarticle}
 \biboptions{comma,sort&compress}
 
\usepackage{graphicx}
\usepackage{here}
\usepackage{ecrc}

\volume{00}

\firstpage{1}

\journalname{Nuclear and Particle Physics Proceedings}

\runauth{}

\jid{nppp}

\jnltitlelogo{Nuclear and Particle Physics Proceedings}

\usepackage{amssymb}

\usepackage[figuresright]{rotating}

\def\nin{\noindent}
\def\beq{\begin{equation}}
\def\eeq{\end{equation}}
\def\bea{\begin{eqnarray}}
\def\eea{\end{eqnarray}}
\def\nnb{\nonumber}
\def\la{\langle}
\def\ra{\rangle}
\def\ga{\left(}
\def\dr{\right)}
\def\lrar              {\Longrightarrow}

\begin{document}

\begin{frontmatter}



\title{$0^+$ and $1^+$ heavy-light exotic mesons at N2LO in the chiral limit}

\author[label]{R.M Albuquerque}
  \address[label]{Faculty of Technology, Rio de Janeiro State University (FAT, UERJ), Brazil }

\ead{raphael.albuquerque@uerj.br}

 \author[label2]{F. Fanomezana}
  \address[label2]{Institute of High Energy Physics of Madagascar (iHEP-MAD), University of Antananarivo, Madagascar }

\ead{fanfenos@yahoo.fr}

 \author[label3]{S. Narison}
  \address[label3]{Laboratoire
Univers et Particules (LUPM), CNRS-IN2P3 \& Universit\'e
de Montpellier II, 
\\
Case 070, Place Eug\`ene
Bataillon, 34095 - Montpellier Cedex 05, France.}
\ead{snarison@yahoo.fr}

 \author[label2]{A. Rabemananjara\corref{cor2}}
\cortext[cor2]{Speaker.}
\ead{{achris$\_$}01@yahoo.fr.}

 \author[label2]{D. Rabetiarivony}
\ead{bidds.davidson@outlook.com}

\author[label2]{G. Randriamanatrika}
\ead{artesgaetan@gmail.com}

\begin{abstract}
\noindent
We use QCD spectral sum rules (QSSR) and the factorization properties of molecule and four-quark currents to estimate  the masses and couplings  of the $0^+$ and $1^{+}$ molecules and four-quark at N2LO of PT QCD. We include in the OPE the contributions of non-perturbative condensates up to dimension-six. Within the Laplace sum rules approach (LSR) and in the $\overline{MS}$-scheme, we summarize our results in Table \ref{tab:results}, which agree within the errors with some of the observed XZ-like molecules or/and four-quark. Couplings of these states to the currents are also extracted. Our results are improvements of the LO ones in the existing literature.

\end{abstract}

\begin{keyword}
QCD Spectral Sum Rules, molecule and four-quark states, heavy quarkonia.


\end{keyword}

\end{frontmatter}


\vspace*{-1cm}
\section{Introduction}
\vspace*{-0.2cm}
\nin
 The recent discovery of the $Z_c$(3900) $1^{++}$ by Belle \cite{BELLE} and BESIII \cite {BES} has motivated different theoretical analysis\,\cite{REVMOL}.
  However, all of the previous analysis like e.g. in \cite{RAPHAEL, HEP1, HEP2} from QCD Spectral Sum Rules (QSSR) \cite{SVZ,SNB1} have been done at LO of PT QCD. In this paper, we are going to use QSSR to evaluate the masses and couplings of some $0^+$ and $1^+$ molecules at N2LO in the PT series and compare the results with those obtained at lowest order and with experiments. This work is a part of the original papers in \cite{ARTICLE} and also in \cite{HEP3, ARTICLE2}. 
\vspace*{-0.25cm}
\section{QCD analysis of the ones in molecule states}
\vspace*{-0.25cm}
\nin
   \subsubsection*{$\bullet$ Currents and two-point fonctions}
   \nin
The currents $J^{(\mu)}_{mol}$ used for these molecules states are given by:\\
 - For $0^+$:
\bea
\bar MM &:& (\bar q \gamma_5 Q)(\bar Q \gamma_5 q)\nnb\\
\bar M^*M^* &:& (\bar q \gamma_\mu Q)(\bar Q \gamma^\mu q)\nnb\\
\bar M^*_0M^*_0 &: &(\bar qQ)(\bar Qq)\nnb\\
\bar M_1M_1 &:& (\bar q \gamma_\mu \gamma_5 Q)(\bar Q \gamma^\mu \gamma_5 q)
\eea
- For $1^+$:
\bea
\hspace*{-.5cm}\bar M^*M &: &\frac{1}{\sqrt{2}}[(\bar Q \gamma_\mu q)(\bar q \gamma_5 Q)-(\bar q \gamma_\mu Q)(\bar Q \gamma_5 q)]\nnb\\
\hspace*{-.5cm}\bar M^*_0M_1 &: &\frac{1}{\sqrt{2}}[(\bar Qq)(\bar q \gamma_\mu\gamma_5 Q)+(\bar qQ)(\bar Q \gamma_\mu\gamma_5 q)],
\eea
where $q$, $Q$ represent respectively light and heavy quarks.
The associated  two-point correlation function is:
\bea
&&\hspace*{-1cm}\Pi^{\mu\nu}_{mol}(q)=i\int d^4x ~e^{iq.x}\la 0
|TJ^\mu_{mol}(x){J^\nu}_{mol}^\dagger(0)
|0\ra
\nnb\\
&&\hspace*{-0.5cm}=-(q^2 g^{\mu\nu}-{q^\mu q^\nu})\Pi_{mol}(q^2)
+q^\mu q^\nu\Pi_{mol}^{(0)}(q^2)~,
\label{2po}
\eea
where $\Pi_{mol}$ and  $\Pi_{mol}^{(0)}$ are associated to the spin 1 and 0 molecule states. 
Parametrizing the spectral function by one resonance plus a QCD continuum, the lowest resonance mass $M_H$ and coupling $f_H$ normalized as:
\beq
\la 0|J^\mu|H\ra=f_H M_H^4 \epsilon^\mu~,
\label{eq:decay}
\eeq
can be extracted by using the Laplace sum rules (LSR)\,\cite{SVZ,BELL,SNR,SNB1,SNB2,SNB3}:
\beq
M^2_H=\frac{\int_{4M^2_Q}^{t_c} dt~ t ~ e^{-t \tau}  \frac{1} {\pi} {\rm Im} \Pi_{mol}(t)}{\int_{4M^2_Q}^{t_c} dt~ e^{-t \tau}  \frac{1} {\pi} {\rm Im} \Pi_{mol}(t)}
\label{mass}
\eeq
and
\beq
f^2_H=\frac{\int_{4M^2_Q}^{t_c} dt~ e^{-t \tau}  \frac{1} {\pi} {\rm Im} \Pi_{mol}(t)}{e^{-\tau M^2_H}M^8_H}
\label{coupling}
\eeq
where $M_Q$ is the heavy quark on-shell mass, $\tau$ the LSR parameter, $t_c$ the continuum threshold and $ {\rm Im} \Pi_{mol}(t)$ is the QCD expression of the molecule spectral function. 
\vspace*{-0.1cm}
 \subsubsection*{$\bullet$ The QCD two-point function at N2LO}
 \nin
To derive the results at N2LO, we assume factorization and then use the fact that the two-point function of a molecule state can be written as a convolution of the spectral functions associated
to quark bilinear currents. We have \cite{CONV,SNPIVO}:
\begin{eqnarray}
\frac{1}{\pi}{\rm Im}\Pi_{mol}^{(0,1)}(t)&=&\theta(t-4M^2_Q)\left(\frac{1}{4\pi}\right)^2 t^2 \int^{(\sqrt{t}-M_Q)^2}_{M_Q^2}\hspace*{-1cm} dt_1 \nnb\\ && \times \int^{(\sqrt{t}-\sqrt{t_1})^2}_{M_Q^2} \hspace*{-1cm} dt_2 ~~~~\times ...
\end{eqnarray}
{ For spin 0:}~~
\begin{eqnarray}
 ...=\lambda^{1/2}
\left[ \left(\frac{t_1}{t}+\frac{t_2}{t}-1\right)^2\right ] \frac{1}{\pi}{\rm Im}\Pi^{(0)}(t_1) \frac{1}{\pi}{\rm Im}\Pi^{(0)}(t_2)\nnb\\
\rm{or}  ~~~...=\lambda^{3/2} \frac{1}{\pi}{\rm Im}\Pi^{(1)}(t_1) \frac{1}{\pi}{\rm Im}\Pi^{(1)}(t_2)\nnb 
\end{eqnarray}
{ For spin 1:}~~
\begin{eqnarray}
 ...=\lambda^{1/2}
\left[ \left(\frac{t_1}{t}+\frac{t_2}{t}-1\right)^2+\frac{8 t_1 t_2}{t^2}\right ]\times \nnb\\
 \frac{1}{\pi}{\rm Im}\Pi^{(0)}(t_1) \frac{1}{\pi}{\rm Im}\Pi^{(1)}(t_2)\nnb
\end{eqnarray}

with the phase space factor:
\bea
\lambda=\left(1-\frac{(\sqrt{t}-\sqrt{t_1})^2}{t}\right)\left(1-\frac{(\sqrt{t_1}+\sqrt{t_2})^2}{t}\right)~.
\eea   
 Im$\Pi^{(1)}(t)$ and Im$\Pi^{(0)}(t)$ are respectively the spectral functions associated to the (axial)vector and   to the (pseudo)scalar bilinear currents.
The QCD expression of the spectral functions for bilinear currents are already known up to order $\alpha_s^2$ and including non-perturbative condensates up to dimension 6. It can be found in \cite{SNFB12,SNFBST14,GENERALIS,CHET} for the on-shell mass $M_Q$.
We shall use the relation between the on-shell $M_Q$ and the running mass $\overline m_Q(\mu)$ to transform the spectral function into the $\overline{MS}$-scheme \cite{SPEC1,SPEC2}: 
\bea
M_Q &=& \overline{m}_Q(\mu)\Bigg{[}
1+\frac{4}{ 3} a_s+ (16.2163 -1.0414 n_l)a_s^2\nnb\\
&&+\ln{\ga\frac{\mu}{M_Q}\dr^2} \ga a_s+(8.8472 -0.3611 n_l) a_s^2\dr\nnb\\
&&+\ln^2{\ga\frac{\mu}{M_Q}\dr^2} \ga 1.7917 -0.0833 n_l\dr a_s^2\Bigg{]},
\label{eq:pole}
\eea
where $n_l=n_f-1$ is the number of light flavours and $a_s(\mu)=\alpha_s(\mu)/\pi$ at the scale $\mu$. 
\subsubsection*{$\bullet$ QCD parameters}
\nin
The PT QCD parameters which appear in this analysis are $\alpha_s$, the charm and bottom quark masses $m_{c,b}$ (the light quark masses have been neglected). 
We also consider non-perturbative condensates  which are the quark condensate $\langle\bar q q\rangle$, the two-gluon condensate $\langle g^2 G^2\rangle$, the mixed condensate $\langle g\bar q G q\rangle$, the four-quark condensate $\rho\langle\bar q q\rangle^2$, the three-gluon condensate $\langle g^3 G^3\rangle$, and the two-quark multiply two-gluon condensate $\rho\alpha_s\langle\bar q q\rangle \langle G^2 \rangle$
  where $\rho$ indicates the deviation from the four-quark vacuum saturation. Their values are given in Table \ref{tab:parameter} and more recently in\,\cite{SN18}.
\vspace*{-0.25cm}
{\scriptsize
\begin{table}[h]
\setlength{\tabcolsep}{.2pc}
 \caption{QCD input parameters:the original errors for $\la\alpha_s G^2\ra$, $\la g^3  G^3\ra$ and $\rho \la \bar qq\ra^2$ have been multiplied by about a factor 3 for a conservative estimate of the errors (see also the text). }  
    {\footnotesize
 {\begin{tabular}{@{}lll@{}}
&\\
\hline
\hline
Parameters&Values& Ref.    \\
\hline
$\alpha_s(M_\tau)$& $0.325(8)$&\cite{SNTAU,BNPa,BNPb,BETHKE}\\
$\hat m_s$&$(0.114\pm0.006)$ GeV &\cite{SNB1,SNTAU,SNmassa,SNmassb,SNmass98a,SNmass98b,SNLIGHT}\\
$\overline{m}_c(m_c)$&$1261(12)$ MeV &average \cite{SNmass02,SNH10a,SNH10b,SNH10c,PDG,IOFFEa,IOFFEb}\\
$\overline{m}_b(m_b)$&$4177(11)$ MeV&average \cite{SNmass02,SNH10a,SNH10b,SNH10c,PDG}\\
$\hat \mu_q$&$(253\pm 6)$ MeV&\cite{SNB1,SNmassa,SNmassb,SNmass98a,SNmass98b,SNLIGHT}\\
$M_0^2$&$(0.8 \pm 0.2)$ GeV$^2$&\cite{JAMI2a,JAMI2b,JAMI2c,HEIDb,HEIDc,SNhl}\\
$\la\alpha_s G^2\ra$& $(7\pm 3)\times 10^{-2}$ GeV$^4$&
\cite{SNTAU,LNT,SNIa,SNIb,YNDU,SNH10a,SNH10b,SNH10c,SNG2,SNGH}\\
$\la g^3  G^3\ra$& $(8.2\pm 2.0)$ GeV$^2\times\la\alpha_s G^2\ra$&
\cite{SNH10a,SNH10b,SNH10c}\\
$\rho \alpha_s\la \bar qq\ra^2$&$(5.8\pm 1.8)\times 10^{-4}$ GeV$^6$&\cite{SNTAU,LNT,JAMI2a,JAMI2b,JAMI2c}\\
\hline\hline
\end{tabular}}
}
\label{tab:parameter}
\vspace*{-0.5cm}
\end{table}
} 
\vspace*{-0.2cm}
\section{Mass of the $ \bar DD(0^{+})$ molecule state}
\vspace*{-0.2cm}
\subsubsection*{$\bullet$ $\tau$ and $t_c$ stabilities}
\begin{figure}[hbt] 
\centerline{\includegraphics[width=7.cm]{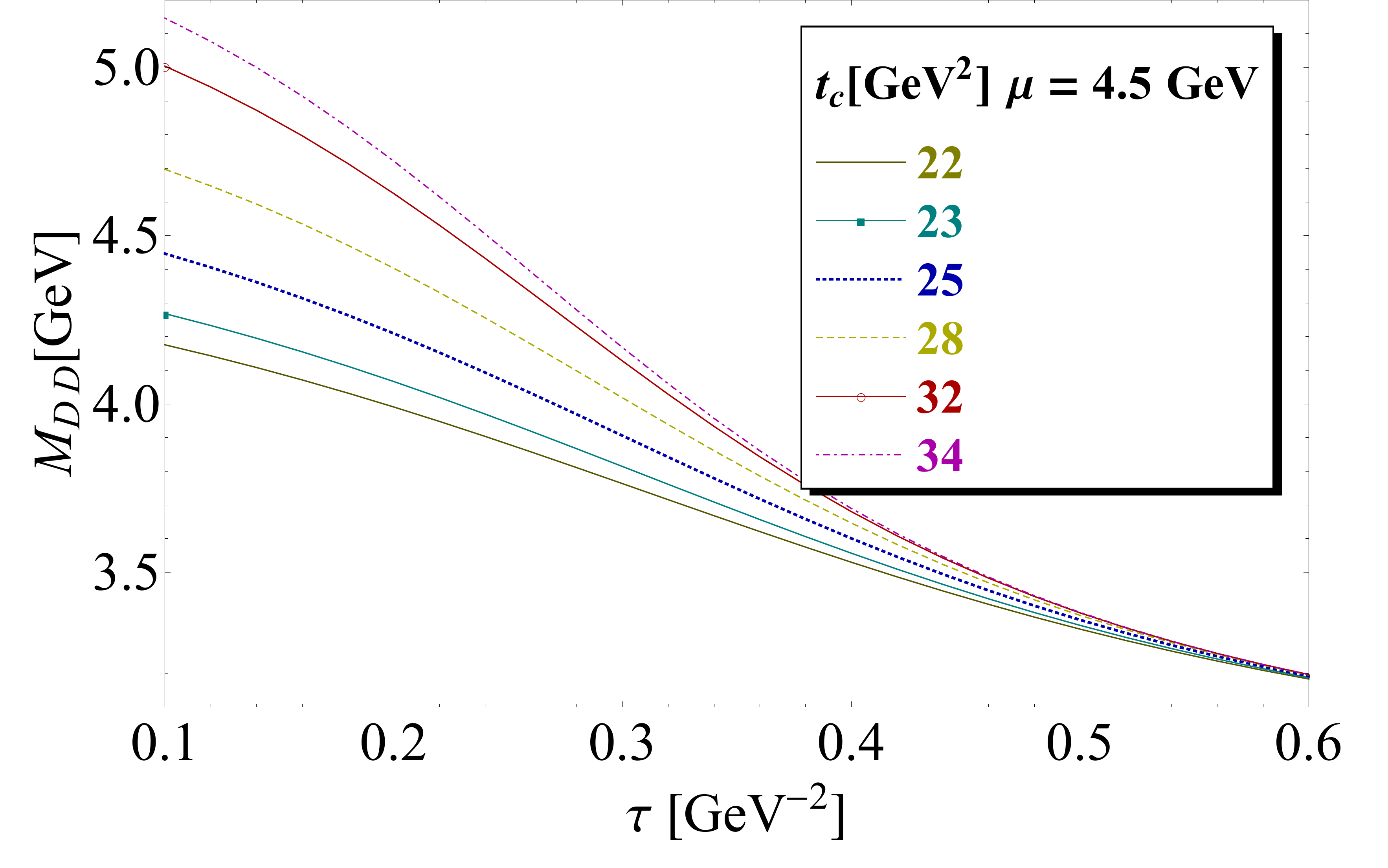}}
\caption{\scriptsize $\tau$-behavior of  $M_{\bar DD}$ at N2LO for different values of $t_c$ and for $\mu$=4.5 GeV}
\label{fig1d} 
\end{figure}

\nin
 We study the behavior of the mass in term of LSR variable $\tau$ at different values of $t_c$ as shown in Fig.\ref{fig1d}. 
 We consider as a final and conservative result the one corresponding to the beginning of the $\tau$ stability for $t_c$=23 GeV$^2$ and $\tau\simeq$ 0.25 GeV$^{-2}$ until the one where $t_c$ stability is reached for $t_c\simeq$ 32 GeV$^2$ and $\tau\simeq$ 0.35 GeV$^{-2}$.
\vspace*{-0.75cm}
\subsubsection*{$\bullet$ Convergence of the PT series} 
\begin{figure}[hbt] 
\centerline{\includegraphics[width=6.cm]{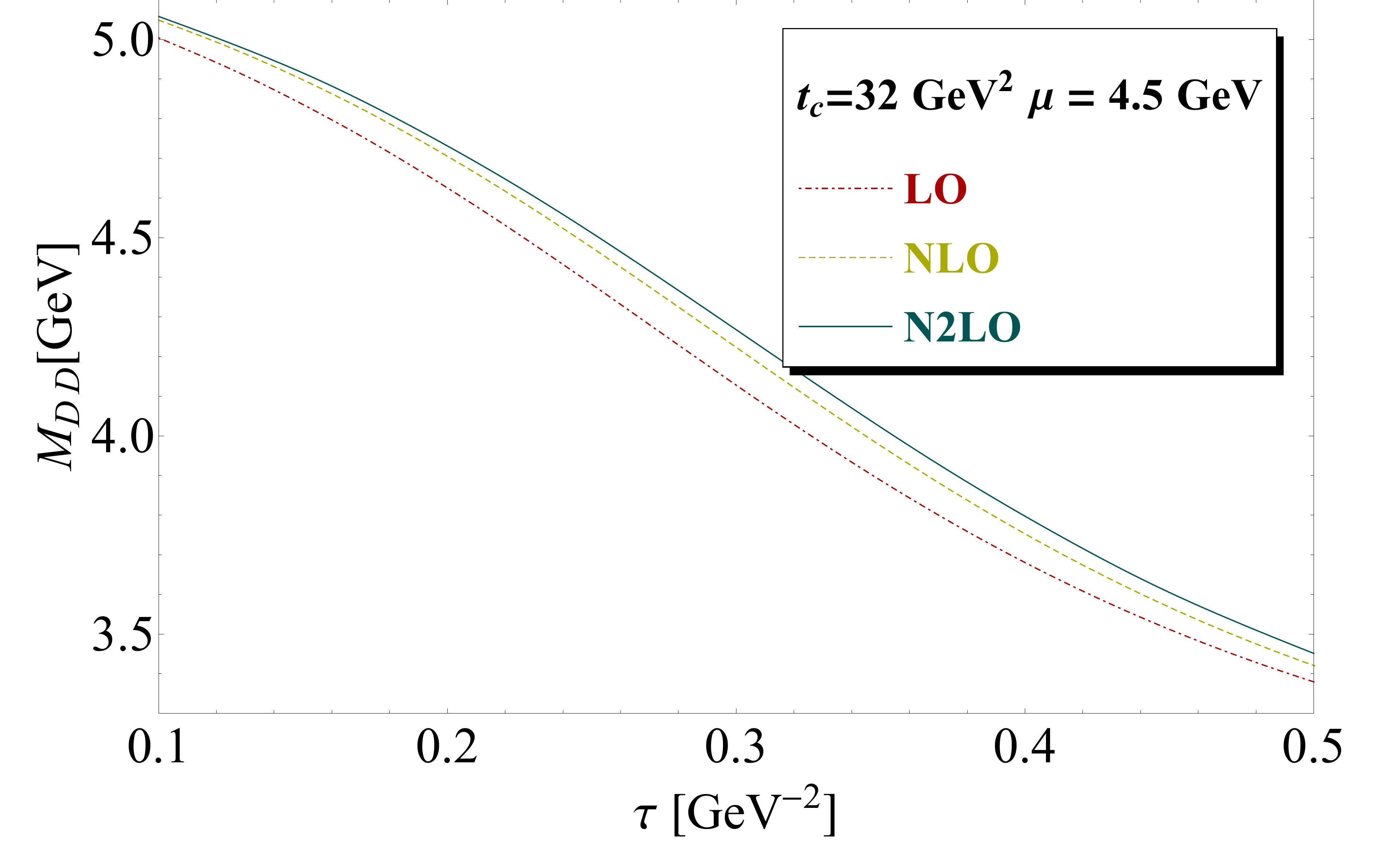}}
\caption{\scriptsize $\tau$-behavior of $M_{\bar DD}$ for $t_c$=32 GeV$^2$ and   $\mu$=4.5 GeV and for different truncation of the PT series}
\label{fig2d} 
\end{figure}

\nin
According to these analysis, we can notice that the $\tau$-stability begins at $t_c=23$ GeV$^2$ and the $t_c$-stability is reached from $t_c= 32$ GeV$^2$. Using these two extremal values of $t_c$, we study in Fig. {\ref{fig2d}} the convergence of the PT series for a given value of $\mu=4.5$ GeV.  We observe that from LO to NLO the mass increases by about +1$\%$ while  from NLO to N2LO, it only increases by +0.1$\%$. This result indicates a good convergence  of PT series which validates the LO result obtained in the literature when the  running quark mass is used \cite{RAPHAEL}.
\subsubsection*{$\bullet$ $\mu$-stability}
\begin{figure}[hbt] 
\centerline{\includegraphics[width=6.cm]{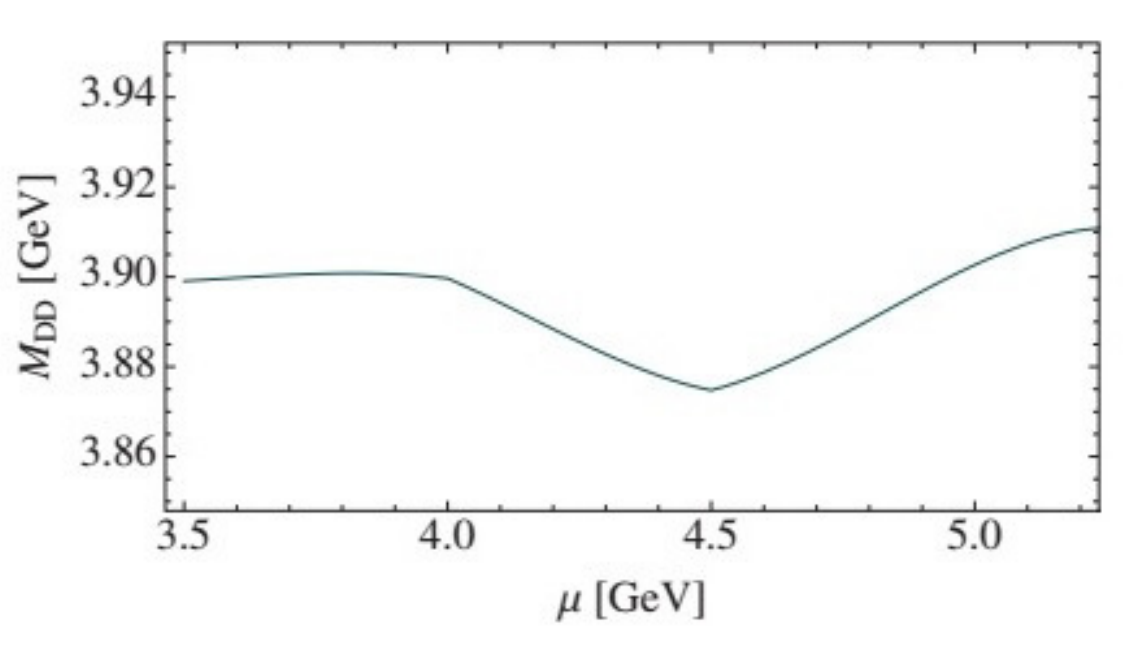}}
\caption{\scriptsize $\mu$-behaviour of $M_{\bar D^*D^*}$  at N2LO}
\label{fig3d} 
\end{figure}

\nin
We improve our previous results by using different values of $\mu$ (Fig. {\ref{fig3d}}). Using the fact that the final result must be independent of the arbitrary parameter $\mu$, we consider as an optimal result the one at the inflexion point
for $\mu\simeq (4.0-4.5)$ GeV: 
\beq
 M_{\bar DD} = 3898(36) \rm{MeV}~,
\label{eq:md*d}
\eeq
where the second error comes from the localisation  of the inflexion point, QCD condensates and higher dimension contributions.
\vspace*{-0.5cm}
\section{Coupling of $ DD(0^{+})$ molecule state}
\begin{figure}[hbt] 
\centerline{\includegraphics[width=6.cm]{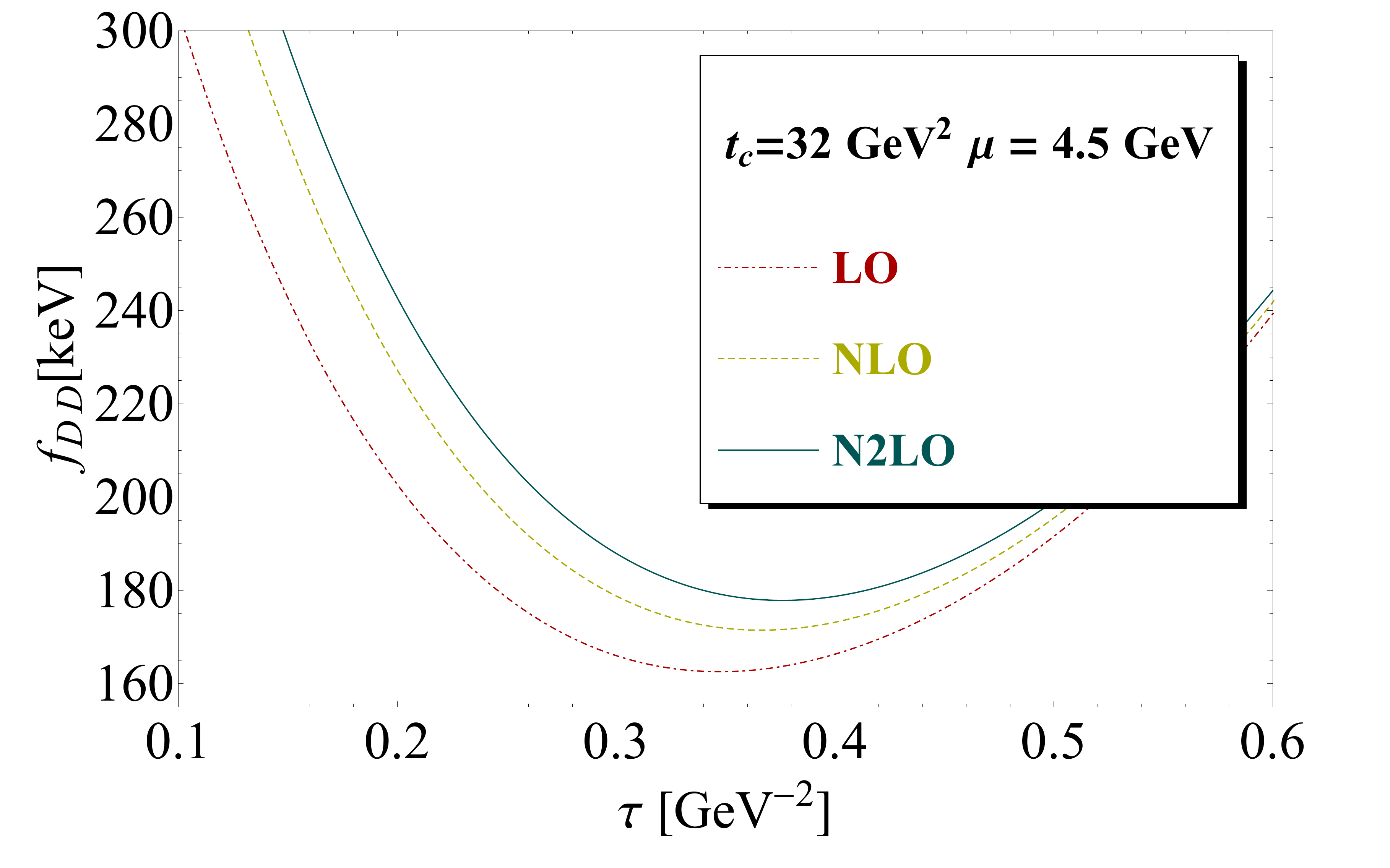}}
\caption{\scriptsize $\tau$-behavior of the running coupling $f_{\bar DD}$ for $t_c$=32 GeV$^2$, $\mu=4.5$ GeV and for different truncation of the PT series}
\label{fig1f} 
\end{figure}

\nin
We can do the same analysis to derive the decay constant $f_H$ defined in Eq. (\ref{eq:decay}). Noting that the bilinear pseudoscalar heavy-light current acquires an anomalous dimension, then the decay constant runs as:
\bea
f^{(s,p)}_H(\mu)&=&\hat f^{(s,p)}_H \ga {-\beta_1 a_s}\dr^{4/\beta_1}/r_m^2~,\nnb \\
f^{(1)}_H(\mu)&=&\hat f^{(1)}_H \ga {-\beta_1 a_s}\dr^{2/\beta_1}/r_m~,
\eea
where $\hat f_H$ is a scale invariant coupling; $−\beta_1 =
(1/2)(11 - 2n_f /3)$ is the first coefficient of the QCD $\beta$-function for $n_f$ flavors and $ a_s \equiv \alpha_s/\pi $. The QCD corrections numerically read:
\bea
r_m(n_f = 4) = 1 + 1.014a_s + 1.389a^2_s, \nnb\\
r_m(n_f = 5) = 1 + 1.176a_s + 1.501a^2_s.
\eea.
Taking the Laplace transform of the correlator, this definition will lead us to the expression of the running coupling in Eq. (\ref{coupling}). We show in Fig. \ref{fig1f} the $\tau$-behaviour of the running coupling $f_{\bar DD}(\mu)$ for two extremal values of $t_c$ where $\tau$ and $t_c$ stabilities are reached. These values  are the same as in the mass determination. 
One can see in this figure that the $\alpha_s$ corrections to the LO term of PT series are still small though bigger than in the case of the mass determination from the ratio of sum rules. It is about +5$\%$ from LO to NLO  and +2$\%$ from NLO to N2LO.  
\begin{figure}[hbt] 
\centerline{\includegraphics[width=6.cm]{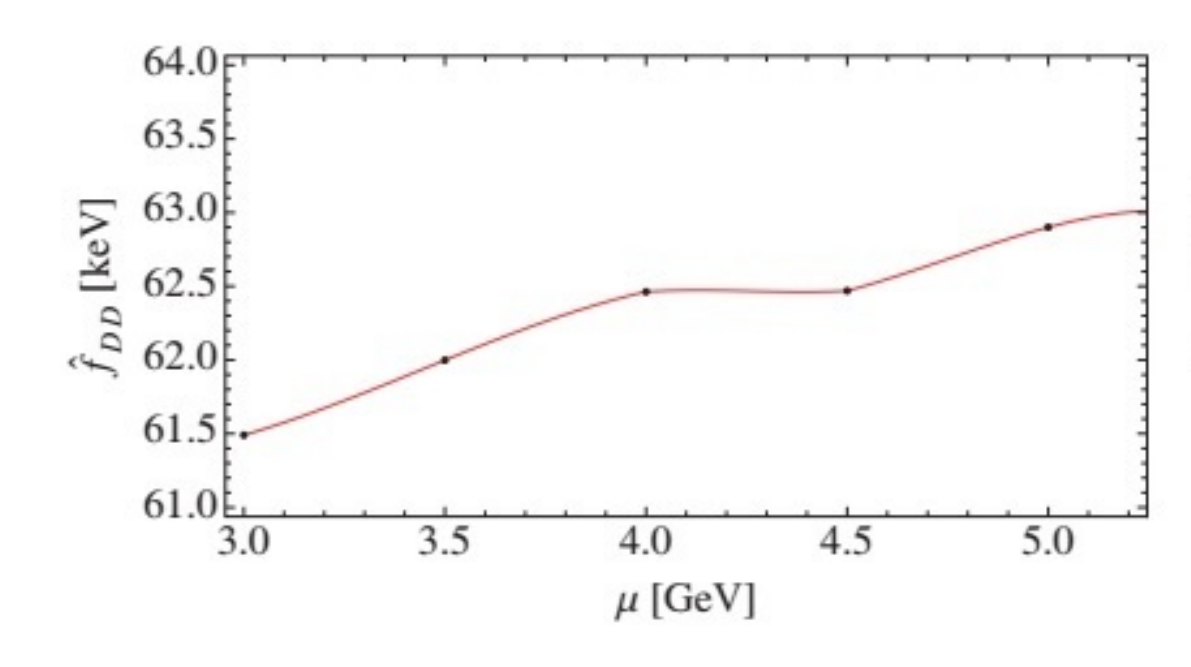}}
\caption{\scriptsize $\mu$-behavior of $f_{\bar DD}$  at N2LO}
\label{fig2f} 
\end{figure} 
In the Fig. {\ref{fig2f}}, we show the $\mu$ behaviour of the invariant coupling $\hat f_{\bar DD}$. Taking the optimal result at the minimum for $\mu\simeq 4.5$ GeV, we obtain in units of MeV:
\beq
\hspace*{-0.25cm}\hat f_{\bar DD}=(62\pm 6)~{\rm keV}\lrar
 f_{\bar DD}(\mu)=(170\pm 15)~{\rm keV},
\label{eq:fd*d}
\eeq
\section{$0^+$ and $1^+ $ four-quark states}
We can do the same analysis for the case of four-quark states. The interpolating currents used  are:
{\footnotesize
\bea 
\hspace*{-1.25cm}&&0^{+}:~ \epsilon_{abc}\epsilon_{dec}\big[\big(q_a^T C \gamma_5 Q_b \big)\big(\bar q_d C \gamma_5 C \bar Q_e^T \big)+ k \big(q_a^T C Q_b \big)\big(\bar q_d C \bar Q_e^T \big)\big]\nnb\\
\hspace*{-1.25cm}&&1^+:~ \epsilon_{abc}\epsilon_{dec}\big[\big(q_a^T C \gamma_5 Q_b \big)\big(\bar q_d C \gamma_\mu C \bar Q_e^T \big) 
+k \big(q_a^T C Q_b \big)\big(\bar q_d \gamma_\mu \gamma_5 C \bar Q_e^T \big)\big],\nnb
\eea
}
where $Q \equiv c$ (respectively $b$) in the charm (resp. bottom) channel and $q \equiv u, d$. $k$ is the mixing of the two operators. We use k=0 as shown in \cite{RAPHAEL, RAPHAEL1}. \\ 
The behavior of the curves of masses and couplings are very similar to the molecules ones. Considering all the possible currents and channels configurations, we have in Table \ref{tab:results} the results for  $0^+$ and $1^+$ molecule and four-quark states.
\section{Conclusions}
\nin
We have presented improved predictions of QSSR for the masses and couplings of the $0^+$ and $1^+$ molecule and four-quark states at N2LO  of PT series and including up to dimension six non-perturbative condensates. We can see a good convergence of the PT series after including higher correction. This good convergence confirms the veracity of our results. 
 The results are improvements of all the precedent works about the masses of exotic hadrons obtained at LO. Our analysis has been done within stability criteria with respect to the LSR variable $\tau$, the QCD continuum threshold $t_c$ and the subtraction constant $\mu$ which have provided successful predictions in different hadronic channels. The optimal values of the masses and couplings have been extracted at the same value of these parameters where the stability appears as an extremum and/or inflection points. The ill-defined heavy quark mass definition used at LO is not enough to have results. The effects are often large for the coupling. The masses of $\bar DD$ and $\bar D^*D^*$ are also around the $Z_c(3900)$ $0^{++}$ state.
We do not include higher dimensioon condensates contributions in our estimate but only consider them as a source of the errors. One can conclude that $Z_c(3900)$ can be
well described with an almost pure $D^*D$ molecule. One can notice that the masses of the $J^P = 1^+, 0^+$ states are most of them below the corresponding $DD$, $BB$-like thresholds and are compatible with some of the observed XYZ masses suggesting that these states can be interpreted as almost pure molecules or/and four-quark states.














\begin{table*}[hbt]
\vspace*{-0.25cm}
\setlength{\tabcolsep}{1.1pc}
\newlength{\digitwidth} \settowidth{\digitwidth}{\rm 0}
\catcode`?=\active \def?{\kern\digitwidth}
\caption{\scriptsize{ $0^+$ and $1^+$ molecules and four-quark masses,  invariant and running couplings from LSR within stability criteria from LO to N2LO of PT.}}
 \scriptsize{ 
 \begin{tabular*}{\textwidth}{@{}l@{\extracolsep{\fill}}ccc ccc ccc}
\hline
\hline
               \bf Channels
                    &\multicolumn{3}{c}{$\hat{f}_X$ \bf [keV]}
					&\multicolumn{3}{c}{$f_X$ \bf [keV]}
					&\multicolumn{3}{c}{$M_X$ \bf [MeV]}\\
\cline{2-4} \cline{5-7} \cline{8-10}
				 & \multicolumn{1}{c}{{LO}}
                 & \multicolumn{1}{c}{{NLO}}
                 & \multicolumn{1}{c}{N2LO}
                 & \multicolumn{1}{c}{{LO}} 
                 & \multicolumn{1}{c}{NLO} 
                 & \multicolumn{1}{c}{N2LO} 
                 & \multicolumn{1}{c}{{LO}} 
                 & \multicolumn{1}{c}{NLO} 
                 & \multicolumn{1}{c}{N2LO} 
                
                  \\
\hline
 \bf{Scalar($0^{+}$)}&&&&&&&&&\\
$\bar DD
$&56&60&62(6)&155&164&170(15)&3901&3901&3898(36)\\
$\bar D^{*}D^{*}$&-&-&-&269&288&302(47)&3901&3903&3903(179)\\
$\bar D^{*}_{0}D^{*}_{0}$&-&-&-&-&97(15)&114(18)&-&4003(227)&3954(223)\\
$\bar D_{1}D_{1}$&-&-&-&-&236(32)&274(37)&-&3858(57)&3784(56) \\
$S_c$&62&67&70(7)&173&184&191(20)&3902&3901&3898(54)\\

\\
$\bar BB
$&4.0&4.4&5(1)&14.4&15.6&17(4)&10605&10598&10595(58)\\
$\bar B^{*}B^{*}$&-&-&-&27&30&32(5)&10626&10646&10647(184)\\
$\bar B^{*}_{0}B^{*}_{0}$&2.1&2.3&4(1)&7.7&11.3&14(4)&10653&10649&10648(113)\\
$\bar B_{1}B_{1}$&-&-&-&-&20(3)&28.6(4)&-&10514(149)&10514(149) \\
$S_b$&4.6&5.0&5.3(1.1)&16&17&19(4)&10652&10653&10654(109)\\

\bf {Axial($1^{+}$)}&&&&&&&&&\\
$\bar D^{*}D$&87&93&97(10)&146&154&161(17)&3901&3901&3903(62)\\
$\bar D^{*}_{0}D_{1}$&-&-&-&-&96(15)&112(17)&-&3849(182)&3854(182)\\
$A_c$&100&106&112(18)&166&176&184(30)&3903&3890&3888(130)\\
\\
$\bar B^{*}B$&7&8&9(3)&14&16&17(5)&10680&10673&10646(150)\\
$\bar B^{*}_{0}B_{1}$&4&6&7(1)&8&11&14(2)&10670&10679&10692(132)\\
$A_b$&8.7&9.5&10(2)&16&18&19(3)&10730&10701&10680(172)\\
\hline
\hline
\end{tabular*}
}
\label{tab:results}
\vspace*{-0.25cm}
\end{table*}
\newpage
\input{bib_sample}

\end{document}